\documentclass[a4paper,10pt]{article}
\usepackage{enumerate}
\usepackage{amsmath}
\usepackage{amsfonts}
\usepackage{amssymb}
\usepackage{amsthm}
\usepackage{graphicx}
\usepackage{geometry}
\usepackage{float}
\usepackage{graphics}
\usepackage{float}
\usepackage{times}
\usepackage{color}
\usepackage{tikz}
\usepackage{fontenc}
\usepackage[english]{babel}
\usepackage{fontenc}
\usepackage[english]{babel}
\theoremstyle{definition}
\newtheorem{theorem}{Theorem}
\newtheorem{definition}{Definition}
\newtheorem{axiom}{Axiom}
\newtheorem{lemma}{Lemma}
\newtheorem{remark}{Remark}
\newtheorem{example}{Example}

\def\G{\mathcal{G}(N)}

\begin{document}
	\title{Weighted position value for Network games}

	\author{Niharika Kakoty$^{1}$, Surajit Borkotokey$^{1*}$, Rajnish Kumar$^{2}$, Abhijit Bora$^{1}$}
	\date{} 
	\maketitle
	
	\begin{center}
		$^1$Department of Mathematics, Dibrugarh University, Dibrugarh, 786004, Assam, India\\
		$^2$Queen's Management School, Queen's University, Belfast BT7 1NN, Northern Ireland, UK\\
		$^*$Corresponding authors' e-mail:sborkotokey@dibru.ac.in
	\end{center}
	
	
	\begin{abstract}		
	 In Network games under cooperative framework, the position value is a link based allocation rule. It is obtained from the Shapley value of an associated cooperative game where the links of the network are considered players. The Shapley value of each of the links is then divided equally among the players who form those links. The inherent assumption is that the value is indifferent to the weights of the players in the network. Depending on how much central a player is in the network, or the ability of making links with other players etc., for example, players can be considered to have weights. Thus, in such situations, dividing the Shapley value equally among the players can be an over-simplistic notion. We propose a generalised version of the position  value: the weighted position value that allocates the Shapley shares proportional to the players' weights. These weights of the players are exogenously given. We provide two axiomatic characterizations of our value. Finally, a  bidding mechanism is formulated to show that any sub-game perfect equilibrium (SPE) of this mechanism coincides with the weighted position value.
	 \end{abstract}
		
		
		
\section{Introduction}
	In recent years, Network games under cooperative framework have been extensively studied to model various socio-economic and political issues, that include businesses related to trading and exchange, collaborations among acquaintances, security issues in the cyber world, sharing of natural resources among stakeholders, problems in artificial intelligence, wireless communication networks, etc., to name a few. In all these situations, the main concern is in the way players are connected to each other (for instance, who talks to whom, who influences whom, who shares information with whom, or which countries have trade agreements with one another, formation of allies in political scenarios) in order to determine the total productivity or value generated by them. The most fundamental question that arises due to this connection is ``how the worth generated by this collective cooperation is allocated among the players?" The distribution of worth among the players in a fair manner has a rich axiomatic history in cooperative game theory.
	\par A co-operative game with transferable utilities or a TU game in short, is a pair $(N,v)$ where $N$ is the player set and $v : 2^N\rightarrow \mathbb{R}$ is a function such that $v(\emptyset)=0$ and is called the characteristic function. It assigns a real number called the worth to each subset of the player set $N$. These subsets are called coalitions. The value $v(S)$ represents the worth of coalition $S$. The task is then to find a suitable allocation of the worth of the grand coalition(i.e., $v(N))$. The most well known and widely used allocation rule is the Shapley value\cite{shapley1953value}. It is the average of the marginal contributions of a player from each of the coalitions that she belongs to.
	\par Myerson\cite{Myerson} introduces the notion of graph restricted games where players in a coalition can generate worth only when they are connected through a network. This is later referred as a \textit{communication situation} by Jackson and Wollinsky\cite{Jackson1996}. They \cite{Jackson1996} extend Myerson's work and propose a class of games where the worth is generated by the networks. Their game is referred to as a \textit{Network game.} Network games are more flexible and richer than communication situations and can be applied to a much wider class of situations. In this paper, our focus is purely on Network games. 
	\par The most challenging problem in Network games is to find a suitable allocation rule that distributes the total worth generated by the network among the players (nodes in the network). In his paper, Myerson\cite{Myerson} proposes an allocation rule for communication situations induced by the Shapley value of TU games. In this framework, only those coalitions are considered for computing the Shapley value in which the players are connected by some links. This allocation rule is called the Myerson value. The extension of the Myerson value to a Network game is studied in \cite{Jackson1996}. Meessen\cite{Meesen} introduces the position value for communication situations. Similar to the Myerson value, the position value is the Shapley value of a TU game associated with the communication situation where links in the network are considered as players and each player is awarded with half of the Shapley payoffs given to the links. The position value is extended to the class of Network games by Slikker\cite{Slikker05a}. 
\par In TU games, the family of weighted Shapley values is first studied in \cite{Kalai,Shapley}. Each weighted Shapley value associates a positive weight with each player. These weights are proportional to their shares in unanimity games. The symmetric Shapley value is the special case where all the weights are the same. Following this idea, we focus on the weighted position value for network games and provide characterizations of this allocation rule that falls in line with the original characterizations of the position value by Slikker and Slikker and van den Nouweland in \cite{Slikker05b, Nouweland}. Motivated by Kalai and Samet, we incorporate the weighted scheme of the position value. A similar study by Owen in \cite{Owen} show that the weighted systems used in \cite{Kalai,Shapley} are a measure of the inputs of the players to form the grand coalition rather than their bargaining abilities. The weighted Shapley values is generalized to communication situations in \cite{Haeringer,slikker2000} and is extended in \cite{Ghintran} to incorporate the negotiation powers of the players in allocating the worth. Our model, however, differ all these in two fronts: first, we consider Network games instead of communication situations and secondly, our weight schemes focus on the weights of the links shared among the players. Since we are giving more importance to the links rather than the players, the introduction of weights to the links is quite natural. To better interpret this idea, we present the following example.
\begin{example}
Consider the co-author model described in \cite{Jackson1996}. The model interprets the nodes as researchers who spend time writing papers. Each node's productivity is a function of its links. A link represents a collaboration between two researchers. The amount of time a researcher spends on any project is inversely proportional to the number of projects that the researcher is involved in. The fundamental utility or productivity of player $i$ given the network $g$ is $$u_i(g)=\sum_{j:ij \in g}w_i(n_i,j,n_j)-c(n_i)$$ where $w_i(n_i,j,n_j)$ is the utility derived by $i$ from a direct contact with $j$ when $i$ and $j$ are involved in $n_i$ and $n_j$ projects respectively and $c(n_i)$ is the
cost to $i$ of maintaining $n_i$ links. We can think of the number of projects $n_i$ and $n_j$ to be the weights of players $i$ and $j$ respectively and the utility derived by each of the players from the ones that have a direct link with them can be attained by calculating their weighted position values.
\newline To be precise, let the weight of player $i$ on the link $ij$ be given by $w_{ij}^i=\dfrac{n_i}{n_i+n_j}$ and the utility derived by $i$ from a direct contact with $j$ be given by $\sum_{j:ij \in g}w_i(n_i,j,n_j)=\sum_{g'\subseteq g}\dfrac{\sum_{j:ij \in g}w_{ij}^i}{|g'|}\lambda_{g'}(v).$
\par One instance of this model can be that of a well-known reputed author or academician who has, in fact, many contacts that represent her links and suppose she at some point of time agrees to collaborate on a project with a new or less experienced researcher. In such a scenario, even though the academician has a higher degree compared to the fresher the latter should be more benefited from their link than the former. Our value captures these issues in a natural and intuitive manner. The position value on the other hand, seems to have failed to model such situations.
\end{example}
\par This paper is organized as follows. In section 2, we present the relevant definitions and notations used to define our model. Section 3 presents two characterizations of the class of weighted position values and in section 4, we provide a bidding mechanism for the class of weighted position values.
\section{Preliminaries}
In this section, we introduce the standard notations and definitions required for the development of our model. Let $N$ be the player set. Denote by small letters $s$, $t$ etc., respectively the size of the coalitions $S$, $T$ etc., of $N$. Also, we simplify the notations by using $S \setminus i$, $S \cup i$ etc., instead of $S \setminus \{i\}$, $S \cup \{i\}$ etc., for $S \subseteq N$. Let $2^N$ denote the set of coalitions of $N$. Let $(N, v)$ be a TU game. When the player set $N$ is fixed, we denote the TU game $(N, v)$ simply by $v$. Let $\G$ denote the class of all TU games with player set $N$.  Let $i,j\in N,\;i\ne j$ be such that $v(S \cup i) = v(S\cup j)$ for all $S \subset N \setminus \{i, j\}$, then $i$ and $j$ are called symmetric players. Let $i\in N$ for the game $(N, v)$ be such that $v(S \cup i) = v(S)$ for each $S\subset N \setminus i$, then $i$ is called a null player in $(N, v)$. A value is an allocation function $\Phi: \G \mapsto \mathbb R^n$ that assigns each member of $\G$ an $n$-dimensional vector $(\Phi_1(v),\cdots, \Phi_n(v))\in \mathbb R^n$. The Shapley value due to \cite{Shapley} denoted by $\Phi^{Sh}: \G \mapsto \mathbb R^n$ is given by 
\begin{equation}\label{eq:shapley1}
\Phi_i^{Sh}(N, v) = \sum_{S \subseteq N\setminus i} \frac{(n-s-1)!s!}{n!}\Big[v(S \cup i)- v(S)\Big],	
\end{equation}
or equivalently,
\begin{equation}\label{eq:shapley}
\Phi_i^{Sh}(N, v) = \sum_{S \subseteq N: i \in S} \frac{\lambda_S(v)}{s}	
\end{equation}
where $\lambda_S(v)$ is called the Harsanyi dividend given by the following formula
\begin{equation*}
\lambda_S(v) = \sum_{T \subseteq S} (-1)^{s-t} v(T).	
\end{equation*}
The Shapley value has several characterizations, among which the one of our interest is seminally given by Shapley himself in \cite{Shapley}. The axioms used in this characterization are as follows. Let $v\in \G$, and $\Phi: \G \mapsto \mathbb R^n$	be a value.
\begin{axiom}
(Efficiency): $\Phi$ satisfies Efficiency i.e., $\sum_{i\in N} \Phi_i(v)= v(N)$.
\end{axiom}
\begin{axiom}
(Symmetry):  $\Phi$ satisfies Symmetry i.e., $\Phi_i(v) = \Phi_j(v)$ for all such $i,j$.	
\end{axiom}
\begin{axiom}
(Linearity): Let $w\in \G$ be another game. $\Phi$ is linear if for $\alpha, \beta \in \mathbb R$, we have $$\Phi(\alpha v + \beta w) = \alpha \Phi(v) + \beta \Phi(w).$$	
\end{axiom}
\begin{axiom}
(Null player) If $i \in N$ be a null player, then $\Phi_i(N, v)= 0$.	
\end{axiom}
\noindent In the following we mention briefly the literature on Network games, borrowed mostly from \cite{borkotokey_rev_1,Caulier,Jackson2005,Slikker05a}.
\begin{itemize}
\item \textit{Networks}
\newline Given the player set $N$ and distinct players $i,j\in N$, a link $ij$  is the pair $\{ i,j \}$ which represents an undirected relationship between $i$ and $j$. Clearly, $ij$ is equivalent to $ji$. 
\par The set of all possible links with the player set $N$ denoted by $g_{N}=\{ij$ $|$ $i,j\in N$, and $i\neq j\}$ is called the complete network.
\par  A network $g$ is a subset of  $g_{N}$. The set of all possible networks on $N$ is $\mathbb G^N=\{ g \mid g\subseteq g_{N} \}$. In particular, every link $l = ij$ for $i,j \in N,\;i \ne j$ in the network $g$ can be considered as a singleton sub-network and therefore, with a slight abuse we use the notation $l \subseteq g$ instead of $l \in g$. The latter notation, we keep for defining a hypothetical player in a link game later.
\par The network $g_0=\emptyset$ is the network without any links, which we refer to as the empty network.
\par Let the number of links in a network $g$ be denoted by $|g|$. Obviously, $l(g_N)= \binom{n}{2} = \tfrac{1}{2} n(n-1)$, and $l(g_0)=0$. 
\par Denote by $g_i=\{ij\in g\mid j\in N\}\subseteq g$ the set of links of player $i$ in $g$. It follows that $|g_i|$ denotes the number of links of player $i$ in $g$.  
\par 

\par For every network $g\in\mathbb G^N$, the neighborhood of every $i \in N$ in $g$ is given by $N(g_{i})=\{j\in N\mid j\neq i$ and $ij\in g\}$. It is the set of players with whom $i$ is directly linked in $g$. Let $n(g_i)=|N(g_{i})|$, the number of players in $g$ that are directly linked to player $i$.
\par The set of players in the network $g$ is denoted by $n(g)$. Thus, $n(g)$ is the total number of players in the network $g$.

\item \textit{Networks on subsets of players}
\newline For $g\subseteq g_{N}$ and $g'\subseteq g$, let $g-g'$ denote the network $g \setminus g'$. Similarly, for $g'\subseteq g_{N} \setminus g$, let $g+g'$ denote the network $g\cup g'$.
\item \textit{Paths}
\newline  A path in $g$ connecting $i$ and $j$ is a set of distinct players $\{i_{1},i_{2},\ldots,i_{p}\}\subseteq N(g)$ with $p\geqslant2$ such that $i_{1}=i$, $i_{p}=j$, and $\{i_{1}i_{2},i_{2}i_{3},\ldots,i_{p-1} i_{p}\}\subseteq g$.
\par We say $i$ and $j$ are connected if a path exists between them and they are disconnected otherwise.
\item \textit{Components}
\newline The network {$h\subseteq g$} is a component of $g$ if for all { $i\in N(h)$} and  { $j\in N(h)$}, $i\neq j$, there exists a path in  {$h$} connecting $i$ and $j$ and for any {$i\in N(h) $} and $j\in N(g)$, $ij\in g$ implies {$ij\in h$}.
\par  Thus, a component is a maximally connected subnetwork of $g$. We denote the set of the components of a network $g$ by $C(g)$.
\item \textit{Isolated players} 
\newline The set of players that are not connected in the network $g$ are collected in the set of isolated players in $g$ denoted by $N_{0}(g)=N\setminus N(g)=\{i\in N\mid N(g_i)=\emptyset\}$. 
\par  Clearly, $N_0 (g_0) = N$.
\item \textit{Value function}\\
A value function on $\mathbb G^N$ is a function $v \colon\mathbb G^N \to \mathbb R$. The set of all possible value functions with the player set $N$ denoted by $\mathbb V^N$ is a $2^{^{\frac{n(n-1)}{2}}}-1$ dimensional vector space. Let $v_{0}$ denote the null value function given by $v_{0}(g)=0$ for all $g \in \mathbb G^N.$ 
\item \textit{Basis for value functions}
\newline An important subclasses of value functions in $\mathbb V^N$ are the class of unanimity value functions, a general member of which is defined as follows.
\begin{itemize}
\item For any network $g \in \mathbb G^N$, the unanimity value function $ u_{g}$ is defined as: 
\[ u_{g}(g')=\begin{cases} 
        1 &  \text{if} ~ g \subseteq g' \\
        0& ~\text{otherwise}.  
    \end{cases}
 \]
 \end{itemize}
This class of value functions forms a standard basis for $\mathbb V^N$.
\item \textit{Component additive value functions}\\
An interesting subclass of value functions is that of the component additive value functions where the value of a given component of a network is independent of the other components of the network. Thus, a value function $v\in \mathbb V^N$ is component additive if 
$$v(g) = \sum_{g' \in C(g)} v(g')\;\;\;\textrm{for any}\;\;g\in \mathbb G^N.$$
The space of the component additive value functions with the player set $N$ is denoted by $\mathbb H^N$, which is a $\sum_{h \in C(g)}(2^{ |h|}-1)$ dimensional subspace of  $\mathbb V^N$. Note that, each $v\in \mathbb H^N$ such that $v \ne v_0$, can be represented as
\begin{equation}\label{eq:1}
v=\sum_{g \in \mathbb G^N}\lambda_{g}(v)u_{g}
\end{equation} 
where $\lambda_{g}(v)$ are the Harsanyi dividends given by 
 $$ \lambda_{g}(v) = \sum_{g'\subseteq g} (-1)^{|g|- |g'|}v(g').$$
Note that in view of corollary 1 in \cite{Nouweland}, $u_g$ is also component additive.
\item \textit{Network games}\\
A Network game is a pair $(N, v)$ where $N$ is the player set, and $v \colon\mathbb G^N \to \mathbb R$ is a component additive value function that satisfies $v (g_0)=0$. \\
We mention here Lemma~1 in \cite{Nouweland} which is used in characterizing the weighted position value at a latter stage.
\begin{lemma}[\cite{Nouweland}, Lemma~1, page 270]\label{lem:1}
Let $v$ be a component additive value function. Then for any $g \in \mathbb G^N$ that has two links which are not in the same component, the coefficients $\lambda_g(v)$ are all equal to $0$.
\end{lemma}
\item \textit{Network allocation rules}
\newline A network allocation rule is a function $Y\colon\mathbb G^N \times \mathbb V^N \to \mathbb R^N$ such that for every $g \in \mathbb G^N$ and every $v \in \mathbb V^N$, $Y_i(g,v)= 0$ whenever $i \in N_0(g)$.
\item \textit{Efficiency}
\newline  An allocation rule $Y$ is efficient if for each $g \in \mathbb G^N$ and $v \in \mathbb V^N$ it holds that $$\sum_{i\in N(g)} Y_i(g, v) = v(g).$$
\par  Thus, an efficient network allocation rule determines how the collective worth generated by a given network $g \in \mathbb G^N$ concerning the Network game $v \in \mathbb V^N$ is allocated to the players.
\item \textit{Additivity}
\newline For each pair of component additive value functions $v, v' \in \mathbb H^N$ and $g\in \mathbb G^N$ if 
$$Y_{i}(g, v + v')=Y_{i}(g, v)+ Y_i(g, v')~~\forall i\in N,$$then $Y$ is said to satisfy \textit{additivity}.
\par This property specifies that there are no externalities in the allocation rule when the players in a network are involved in several situations, each described by their value functions.
\item \textit{Superfluous link}
\newline A link $l \subseteq g$ is \textit{superfluous} in a Network game $(g,v)$ if
$$v(g')=v(g'\setminus l)$$ for all $g'\subseteq g$ such that $l \subseteq g'$. 
\par A superfluous link does not affect any network it is involved in.
\item \textit{Superfluous Link Property}
\newline An allocation rule $Y$ on $\mathbb G^{N}\times \mathbb V^{N}$ satisfies the \textit{Superfluous Link Property} if
$$Y(g,v)=Y(g\setminus l,v)$$ for all Network game $(g,v)$ in $\mathbb{G}^N\times \mathbb{V}^N$ and all links $l$ that are superfluous in $(g,v)$.
\par The Superfluous Link Property states that the presence or absence of a link does not affect the value of any network as well as the players' allocations in a network.
\item \textit{Link anonymous value function}
\newline A value function $v \in \mathbb{V}^N$ is \textit{link anonymous} on $g$ if $$v(g')=v(g'')$$ for all subnetworks $g',g''\subseteq g$ such that $|g'|=|g''|.$
\item \textit{Link Anonymity}
\newline An allocation rule $Y$ on $\mathbb{G}^N\times \mathbb{V}^N$ is \textit{link anonymous} if for every network $g\in \mathbb{G}^N$ and value function $v \in \mathbb{V}^N$ that is link anonymous on $g$, there exists an $\alpha \in \mathbb R$ such that
$$Y_i(g,v)=\alpha |g_i|$$ for all $i \in N$.
\par Link Anonymity states that when all the links in a network are interchangeable to determine the values of subnetworks, the relative allocations of the players in the network are determined by the relative number of links that each player is involved in.
\item \textit{The  position value}
\newline The position value is the network allocation rule according to which each player $i \in N$ in a network receives half of the Shapley value of each of her links from the corresponding link game.  More formally, the position value is denoted by $Y \colon \mathbb G^ N\times \mathbb V^N \to \mathbb R^N$ and defined by
 \begin{equation}\label{eq:pos}
 Y_{i}(g,v)= \sum_{g'\subseteq g}\frac{1}{2}\dfrac{l(g'_i)}{|g'|}\lambda_{g'}(v)=\sum_{g' \subseteq g}\sum_{l \subseteq g_i}\frac{\lambda_{g'}(v)}{2 |g'|}
 \end{equation}
 for every $g \in \mathbb G^N$ and $v \in \mathbb V^N$.\\
 It follows from Eq.(\ref{eq:pos}) that, given $g \in \mathbb G^N$ and $v \in \mathbb V^N$, the position value of player $i$ from the network game $(N, v)$ is calculated by considering the Shapley values allocated to the links connected to player $i$. These links are treated as individual players and the worth of this game is derived from the  game $v$ treated as a TU game.
 The value $Y_{i}(g,v)$ is then obtained by summing up the Shapley values equally divided among the players (links) that are connected to player $i$. This captures the overall contribution of player $i$'s links to her position within the network game.
 \end{itemize}
 Among many characterizations of the position value for Network games, the following characterization theorem due to \cite{Nouweland} is important for us for the development of our model of the weighted position value.
\begin{theorem}[\cite{Nouweland}, page 270, theorem 2]\label{them:position}
\textit{The position value is the unique allocation rule on $G^ N\times \mathbb V^N$ that is efficient and additive and satisfies the Superfluous Link Property and Link Anonymity.}
\end{theorem} 
\section{The Weighted position value}
 As an a-priori requirement to the definition of the weighted position value, we define the weight system for the class of network games. We assume that each player is endowed with a weight. A weight function is a one-one mapping from the player set to the set of real numbers, i.e., $w \colon N \to \mathbb R^+\cup \{0\}$ such that $w(i) = w_i$ is the weight associated with player $i$. We define the weight of player $i$ due to its link $l:=\{ij\}$ ($j \in N$) by $w^{i}_l=\dfrac{w_i}{w_i+w_j}$ whenever $l \ne g_0$ and $w^{i}_l=0$ for $l = g_0$.  Thus, for each link $l \subseteq g \ne g_0$, we have $0\leq w^i_l\leq 1$. Denote by $\mathbb W^N$ the class of all possible weight distributions over $N$. We represent an element of $\mathbb W^N$ by the weight function $w$. Without loss of generality, we assume that for $g \ne g_0$, there is at least one $l= ij \in g$ such that $w_l^i \ne 0$ and $w_l^j \ne 0$.\\
 For a given $w \in \mathbb W^N$,  the weighted position value $Y^{w}\colon \mathbb G^N \times \mathbb V^N \to \mathbb R^N$ is the allocation rule according to which each player $i \in N$ in a network game $(N,v)$ gets her payoff according to the following formula.
	\begin{align}\label{eq:wp1}
		Y_i^w(g, v)=\sum\limits_{l\in g_i } w_l^i\sum_{l \in g'\subseteq g}	\frac{\lambda_{g'}(v)}{|g'|} .
	\end{align}
 Note that the classical position value of player $i$ given by Eq.(\ref{eq:pos}) can be computed as the sum of the Shapley values evenly distributed among the players associated with her links. On the other hand, for the weighted position value given by Eq.(\ref{eq:wp1}), the distribution of the Shapley value among the players in a link is proportional to the weights of the players. However, if we consider $w_i = $ constant for all $i \in N$, then, the weighted position value is identical with the position value.
In the next section, we provide two axiomatic characterizations of the weighted position value.
\subsection{Characterization}
In this subsection, we provide the axioms to be used in the characterization of the weighted position value; some of which are standard in cooperative game theory and are trivial extensions to their network counterparts, and some are specific to Network games. The first lemma shows that the weighted position value satisfies is Efficiency. 
	\begin{lemma}\label{lemma1}
	The weighted position value satisfies Efficiency, namely, for all $g \in \mathbb G_N$, $w\in \mathbb W^N$, and $(N, v)\in \mathbb V^N$, we have
	\begin{equation*}
		\sum_{i\in N} Y^w_i(g, v) = v(g).
	\end{equation*}
\end{lemma}
	\begin{proof}
		Let $v\in \mathbb{V^{N}}$, $g\in \mathbb{G^{N}}$, and $w \in \mathbb W^N$. Then using the property that $w^{i}_{ij}+w^{j}_{ij}=1$, we have
		\begin{align*}
		\sum_{i\in N} Y_i^w(g,v) &=\sum_{i\in N}\sum_{l\in g_i} w_l^{i}\sum_{l \in g' \subseteq g}\frac{\lambda_{g'}(v)}{|g'|}\\
		&=\sum_{i \in N}\sum_{j \in N(g_i)} w^{i}_{ij}\Big( \sum_{l \in g' \subseteq g}\frac{\lambda_{g'}(v)}{|g'|}\Big) \\
		&=\sum_{i,j \in N}\Big(w^{i}_{ij}+w^{j}_{ij}\Big) \sum_{ij \in g' \subseteq g}\frac{\lambda_{g'}(v)}{|g'|}\\
	    &= \sum_{l \in g' \subseteq g}\frac{\lambda_{g'}(v)}{|g'|}\\
		&=v(g).
		\end{align*}
		This completes the proof.
	\end{proof}
\noindent We now show that the weighted position value satisfies \textit{Additivity.}
	\begin{lemma}\label{lemma2}
		The weighted position value satisfies \textit{Additivity}.
\end{lemma}
\noindent The Additivity property follows directly from the fact that the unanimity coefficients $\lambda_{g'}(\cdot)$ in Eq.(\ref{eq:wp1}) satisfy the additivity property, namely, $\lambda_{g'}(\alpha v + \beta u) = \alpha \lambda_{g'}(v) + \beta \lambda_{g'}(u)$ for $u, v \in \mathbb V^N$ and $\alpha, \beta \in \mathbb R$.\\
The next lemma shows that the weighted position value satisfies \textit{Superfluous Link Property.} .
\begin{lemma}\label{lemma3}
			The weighted position value satisfies \textit{Superfluous Link Property}.
\end{lemma}
\noindent The proof of lemma~\ref{lemma3} follows exactly the same line of reasoning as that of Lemma~5 in \cite{Nouweland}, and therefore, we omit the proof here.	\\
In the next axiom, which we call the \textit{Weighted Link Anonymity}, we depart from the standard axiom of \textit{Link Anonymity} of the position value, in the sense that unlike the number of links of any player being accounted in case of Link Anonymity, we use here, her weights along these links. 
\begin{axiom}\rm \textbf{Weighted Link Anonymity:}\label{axiom4}	
 An allocation rule $Y$ on  $\mathbb G^N  \times \mathbb V^N$ satisfies \textit{Weighted Link Anonymity} if for every $g\in \mathbb G^N$, $v\in \mathbb V^N$ that is \textit{link anonymous} on $g$, and $w \in \mathbb W^N$, there exists an $\alpha \in \mathbb{R}$ such that $$Y_i(g,v)=\alpha \sum\limits_{l\in g_i } w_l^i  \;\textrm{for all $i \in N$.}$$

\end{axiom}
	\begin{lemma}\label{lemma4}
	The weighted position value satisfies \textit{Weighted Link Anonymity}.
	\end{lemma}
	\begin{proof} Let $g \in \mathbb G^N$ and $v \in \mathbb V^N$ be link anonymous in $g$. Then, it is shown in Lemma 6 in \cite{Nouweland} that $\lambda_{g'}(v) = \lambda_{g''}(v)$ for all pairs $g', g'' \subseteq g$ with $|g'| = |g''|$. It follows that there exists an $\alpha \in \mathbb R$ such that 
	$$\sum_{g' \subseteq g : l \in g'} \frac{\lambda_{g'}(v)}{|g'|} = \alpha$$
	Hence, we obtain,
	$
		Y_i^w(g,v) = \displaystyle\sum_{l \subseteq g_i}w^i_l \sum_{g' \subseteq g : l \in g'} \frac{\lambda_{g'}(v)}{|g'|}=\alpha\sum_{l \subseteq g_i}w^i_l
	$
\end{proof}
\begin{remark}
If the weights on each of the players are same, then the \textit{Weighted Link Anonymity} coincides with \textit{Link Anonymity}.
\end{remark}
The following theorem gives the first characterization of the weighted position value.
\begin{theorem}
	The weighted position value $Y^{w}$ is the unique allocation rule on $\mathbb G^N \times \mathbb V^N$ that is \textit{Efficient}, \textit{Additive}, satisfies the  \textit{Superfluous Link Property} and \textit{Weighted Link Anonymity}.
\end{theorem}
\begin{proof} Lemma \ref{lemma1}, \ref{lemma2}, \ref{lemma3} and \ref{lemma4} shows that the weighted position value satisfies  \textit{Efficiency}, \textit{Additivity}, \textit{Superfluous Link Property} and \textit{Weighted Link Anonymity}. 
\\We now show that it is uniquely defined by these axioms. We follow the proof of theorem 2 in \cite{Nouweland} and proceed exactly in the same manner, however for the completeness of the proof we detail the procedure.
	Let $g\in \mathbb{G}^N$, $v\in \mathbb{V}^N$ and $w \in \mathbb W^N$.
Let $Y: \mathbb G^N \times \mathbb V^N \mapsto \mathbb R^N$ be an allocation rule that satisfies the above axioms.
	Recall $v=\sum\limits_{\substack{g'\in \mathbb G^N\\ g'\neq g_0}} \lambda_{g'}(v)u_{g'}$. By \textit{Additivity},
	
	$$ Y^w (g,v)=\sum\limits_{g'\subseteq g} Y^w(g, \lambda_{g'}(v)u_{g'}).$$
	
	Therefore, it is sufficient to show that $Y^w(g,\lambda_{g'}(v)u_{g'})$ is uniquely determined for all $g'\in \mathbb G^N$. For $g'\in \mathbb G^N$ let $v_{g'}=\lambda_{g'}(v)u_{g'}.$
	The following two cases arise:\\
	\noindent\textbf{Case 1:}
		Suppose $g'\setminus g \ne g_0$. \\
		Then $v_{g'}(g'')=0$ for all $g''\subseteq g$. In particular, $v_{g'}(g)=0$.
		Therefore, $v_{g'}$ is \textit{link anonymous} on $g$ and by \textit{Weighted Link Anonymity} of $Y^w$, there exists an $\alpha\in \mathbb{R}$ such that 
	\begin{equation}\label{eq:6}
		Y_i^w(g, v_{g'})=\alpha \sum_{l\in g_i} w_l^i.
	\end{equation}
	\noindent Let $g = g_0$. Then, $Y_i^w(g_0, v_{g'})=0$ for all $i\in N$. \\
	For $g\neq g_0$, using \textit{Efficiency} of $Y^w$, we have $\sum_{i\in N} Y_i^w(g,v_{g'})= v_{g'}(g)=0$.
Invoking Eq.(\ref{eq:6}), we get
\begin{eqnarray*}
	\sum_{i \in N} \alpha (\sum_{l \in g_i} w_l^i) = 0\Rightarrow \alpha \sum_{i \in N}(\sum_{l \in g_i} w_l^i) = 0 \Rightarrow \alpha =0 \Rightarrow Y_i^w(g, v_{g'})=0; \forall i \in N.
\end{eqnarray*}
The last equality follows from the fact that $\displaystyle\sum_{i \in N}\big(\sum_{l \in g_i} w_l^i\big)>0$ as $ 0 \leq w_l^i \leq 1$ for all $i \in N$ with at least one $l \in g$ such that the conditions $w_l^i >0$ and $w_l^j>0$ are satisfied.\\
\noindent\textbf{Case 2:} Suppose $g'\subseteq g$. Let $l\in g\backslash g'$ and $g''\subseteq g$ with $l\in g''$. Then $v_{g'}(g) = \lambda_{g'}(v)$. Moreover,
		\begin{align*}
			v_{g'}(g'')&=v_{g'}(g''\backslash l)\\
					&=\begin{cases}
						\lambda_{g'}(v), \textit{ if } g'\subseteq g''\\
						~~~~0, ~otherwise.
					\end{cases}
		\end{align*}
	Thus, in either case, every link $l\in g\backslash g'$ is \textit{superfluous} in $(g,v_{g'})$. Thus, by the \textit{Superfluous Link Property}, we have
	\begin{equation}\label{eq:7} 
		Y^w(g, v_{g'}) = Y^w(g \setminus l, v_{g'})
	\end{equation}
Repeated application of the \textit{Superfluous Link Property} on Eq.(\ref{eq:7}) will finally give us,
\begin{equation}\label{eq:7} 
		Y^w(g, v_{g'}) = Y^w(g', v_{g'})
	\end{equation}
By construction, $v_{g'}$ is \textit{link anonymous} on $g$, so by the \textit{Weighted Link Anonymity}, there exists an $\alpha\in\mathbb{R}$ such that for each $i\in N$, it holds that
	\begin{equation}\label{eq2}
		Y_i^w(g', v_{g'})= \alpha \sum_{l\in g'_i} w_l^i.
	\end{equation}
	Using Efficiency, we have
\begin{eqnarray*}
\lambda_{g'}(v) &=& v_{g'}(g)= \sum_{i \in N} Y_i^w(g, v_{g'})\\
	&=& \sum_{i \in N}\alpha \sum_{l \in g_i} w_l^i\\
	&=& \alpha \sum_{\substack{i,j\in N \\ ij\in g_i \cap g_j}} \Big(w_{ij}^i +w_{ij}^j\Big) = \frac{\alpha}{2}|g|.
\end{eqnarray*}
It follows that the value $Y(g, v_{g'})$ is uniquely determined for each $g' \in \mathbb G^N$.
\end{proof}
\noindent For the second characterization, we need the following two axioms, namely, the \textit{Component Balancedness} and \textit{Weighted Balanced Link Contributions Property}.

\begin{axiom}\rm \textbf{Component balance}:  \label{axiom1}	 An allocation rule $Y$ is \textit{component balanced} if for every $v \in \mathbb H^N$ and $g \in \mathbb G^N$ it holds that $$\sum_{i\in N(h)}Y_{i}(g,v)=v(h), \;\textrm{for every component $h \in C(g)$.}$$
\end{axiom}
\begin{axiom}\rm \textbf{Weighted Balanced Link Contributions Property:}\label{axiom5}
For every $g\in \mathbb G^N$, $v\in \mathbb H^N$, every pair $i,j \in N$, and $w \in \mathbb W^N$,  an allocation rule $Y$ satisfies the \textit{Weighted Balanced Link Contributions Property} if
$$\sum\limits_{l\in g_j } w_l^j \Big(Y_i(g,v)-Y_i(g\backslash l,v)\Big)=\sum\limits_{l\in g_i } w_l^i \Big(Y_j(g,v)-Y_j(g\backslash l,v)\Big).$$
\end{axiom}
\noindent Here, we are allowing different weights for the participating players in a link. If the weights are taken as identical, then this axiom coincides with the \textit{Balanced Link Contributions Property}\footnote{In his paper, Slikker \cite{Slikker05a} showed that the \textit{Balanced Link Contributions Property} and \textit{Component Balancedness} together characterizes the  position value in communication situations.} due to \cite{Slikker05a}. 
\begin{lemma}\label{lemma:6}
The weighted position value is component balanced.	
\end{lemma}
\begin{proof}
Let $v \in \mathbb H^N$ be given. Let $g \in \mathbb G^N$ and $h \in C(g)$. It follows from Eq.(\ref{eq:1}) that $v$ is a unique linear combination of the value functions $u_{g'}$ for $g_0 \ne g' \subseteq g$ given by 
\begin{equation}\label{eq:10}
	v = \sum_{g_0 \ne g' \subseteq g} \lambda_{g'}(v) u_{g'}
\end{equation}
Note that $u_{g'}$ for each $g'$ is also component additive. Now, by Additivity of the weighted position value, we have
$$Y_i^w(g, v) = \sum_{g_0 \ne g' \in \mathbb G^N} Y^w_i(g, \lambda_{g'}(v) u_{g'}) = \sum_{g_0 \ne g' \in \mathbb G^N} Y^w_i(g, v_{g'}).$$
where $v_{g'} = \lambda_{g'}(v) u_{g'}$. Thus, it is sufficient to show component balancedness of $Y^w(g, v_{g'})$.\\
Suppose $g' \setminus h \ne g_0$. Then, $h$ being a maximally connected network, $g'$ will have at least two links, which are not in the same component. Then, by lemma~\ref{lem:1}, we have $\lambda_{g'}(v) =0$ and therefore, $v_{g'} =0$. It follows that $Y^w_i(g, v_{g'})=0$ for all $i \in N$. Thus, in particular, $Y^w_i(g, v_{g'})=0$  for all $i \in N(h)$. We have $\displaystyle\sum_{i \in N(h)} Y_i(g, v_{g'})=0 = v_{g'}(h)$.\\ 
Next, suppose $g' \subseteq h$. Then $v_{g'}(h)=1$. It is not difficult to see that for all $i \in N \setminus N(h)$, $Y^w_i(g, v_{g'})=0$. By Efficiency, we have,
\begin{eqnarray*}
	\sum_{i \in N} Y^w_i(g, v_{g'}) = v_{g'}(g) = \lambda_{g'}(v)~~~~~~~~~~~~~~~~~~~~~~~~~~~~~~~~~~~~~~~~~~~~~~~~~~~~~\\
	\Rightarrow  \sum_{i \in N \setminus N(h)} Y^w_i(g, v_{g'}) + \sum_{i \in N(h)} Y^w_i(g, v_{g'})~~~~~~~~~~~~~~~~~~~~~~~~~~~~~~~~~~~~\\ 
	\Rightarrow  \sum_{i \in N(h)} Y^w_i(g, v_{g'}) = \lambda_{g'}(v) u_{g'}(h) = v_{g'}(h).~~~~~~~~~~~~~~~~~~~~~~~~~~~~~~~
\end{eqnarray*}
This completes the proof.
\end{proof}
\begin{lemma}\label{lemma5}
	The weighted position value satisfies \textit{Weighted Balanced Link Contributions Property}.
\end{lemma}
\begin{proof} 
Let $i, j\in N(g)$ such that $ i \ne j$. It follows that
	\begin{align*}
		\sum\limits_{l\in g_j } w_l^j \Big(Y_i^w(g,v)-Y_i^w(g\backslash l,v)\Big) &= \sum_{l\in g_j}w_l^j \Bigg(\sum_{k\in g_i}w_k^i \sum_{g' \subseteq g: k \in g'} \frac{\lambda_{g'}(v)}{|g'|}-\sum_{k\in (g\setminus l)_i}w_k^i \sum_{g' \subseteq g\setminus l : k \in g'} \frac{\lambda_{g'}(v)}{|g'|}\Bigg)\\
		    &= \sum_{l \in g_j}\sum_{l \in g_i} w_l^j w_l^i \sum_{g' \subseteq g: l \in g'} \Bigg(\frac{\lambda_{g'}(v)}{|g'|}\Bigg)\\
		    &= \sum\limits_{l\in g_i } w_l^i \Big(Y_j^w(g,v)-Y_j^w(g\backslash l,v)\Big).
	\end{align*}
	This completes the proof.
\end{proof} 
\begin{theorem}\label{theorem 3}
	The weighted position value is the unique allocation rule on $\mathbb{G}^N\times \mathbb{H}^N $ that is \textit{\textit{component balanced}} and satisfies \textit{Weighted Balanced Link Contributions Property}.
\end{theorem}
\begin{proof}
It is clear from lemma \ref{lemma:6} and \ref{lemma5} that the weighted position value is \textit{component balanced} and satisfies the \textit{Weighted Balanced Link Contributions Property}. 
 Now, we show that the weighted position value is the unique allocation rule on the class of component additive network games that satisfies \textit{\textit{component balancedness}} and the \textit{Weighted Balanced Link Contributions Property\footnote{A similar approach to the proof of uniqueness can be found in \cite{Jorzik}.}}.
\\Suppose  $Y^{w}$ and $Y$ be two allocation rules that satisfy both these axioms. Consider a network $g\neq g_0$ with a minimal number of links such that for all $i \in N(g)$, $Y^{w}(g,v)\neq Y(g,v)$. 
Then, for any $l \subseteq g$ and $i \in N(g)$ we have 
\begin{equation}\label{eq9}
Y^{w}(g\setminus l,v)=Y(g\setminus l,v).
\end{equation} 
 Without loss of generality, we can assume that $g$ is connected. Given $i,j \in N(g)$, $i \ne j$, we have from Eq.(\ref{eq9}) and the \textit{Weighted Balanced Link Contributions Property}
 \begin{eqnarray*}
 \sum_{l \subseteq g_j}w^{j}_l\Big[Y_i(g,v)-Y_i(g\setminus l,v)\Big]=\sum_{l \subseteq g_i}w^{i}_l\Big[Y_j(g,v)-Y_j(g\setminus l,v)\Big]~~~~~~~~~~~~~~~~~~~~~~~~~~~~~~~~~~\\
 \implies  \sum_{l \subseteq g_j}w^{j}_l Y_i(g,v)-\sum_{l \subseteq g_i}w^{j}_lY_j(g,v) = \sum_{l \subseteq g_j}w^{j}_l Y_i(g\setminus l,v)-\sum_{l \subseteq g_i}w^{j}_l Y_j(g\setminus l,v)~~~~~~~~\\
 \implies  \sum_{l \subseteq g_j}w^{j}_l Y_i(g,v)-\sum_{l \subseteq g_i}w^{j}_lY_j(g,v) = \sum_{l \subseteq g_j}w^{j}_l Y^w_i(g\setminus l,v)-\sum_{l \subseteq g_i}w^{j}_l Y^w_j(g\setminus l,v)~~~~\\
 \implies  \sum_{l \subseteq g_j}w^{j}_l Y_i(g,v)-\sum_{l \subseteq g_i}w^{j}_lY_j(g,v) = \sum_{l \subseteq g_j}w^{j}_l Y^w_i(g,v)-\sum_{l \subseteq g_i}w^{j}_l Y^w_j(g,v)~~~~~~~~~~~~~~\\
   \end{eqnarray*}
After re-arranging the terms from both the sides of the above expression, we get
\begin{equation}\label{eq:12}
\sum_{l \subseteq g_j}w^{j}_l \Bigg[Y_i(g,v)-Y_i^w(g,v)\Bigg] = \sum_{l \subseteq g_i}w^{i}_l \Bigg[Y_j(g,v)- Y^w_j(g,v)\Bigg].
\end{equation}
Now, Eq.(\ref{eq:12}) holds for every pair of players $i,j \in N(g)$. From Component Efficiency of $Y$ and $Y^w$, we get,
\begin{equation}\label{eq:13}
\sum_{i \in N(g)} Y_i(g, v) = v(g) = \sum_{i \in N(g)} Y_i^w(g, v).	
\end{equation}
Now, taking $Z_i = Y_i(g, v) - Y_i^w(g, v)$ for all $i \in N(g)$, we get a system of $n(g)$ homogeneous linear equations given by Eq.(\ref{eq:12}) and Eq.(\ref{eq:13}) in $n(g)$ variables. Thus, this system has a unique solution, namely, $Z_i = Y_i(g, v) - Y_i^w(g, v) = 0$ for all $i \in N$. This completes the proof.
\end{proof}
\section{A bidding mechanism}\label{sec:5}
In this section, we study the non-cooperative foundation of the weighted position value following the work of ~\cite{slikker07}. This non-cooperative foundation gives us a strategic approach to characterize the value. In this approach, the payoff of a cooperative value arises as a result of players' equilibrium behavior through a bidding mechanism. In his paper \cite{slikker07}, Slikker proposes three bidding mechanisms that implement the Myerson value, the position value, and the component-wise egalitarian value under sub-game perfect Nash equilibrium, respectively. Inspired by this work, we propose our bidding mechanism that implements the weighted position value under sub-game perfect Nash equilibrium. Throughout the bidding mechanism, we assume that the network $g \in \mathbb G^N$ and the underlying Network game $v \in \mathbb H^N$ are known to us. The mechanism is recursive and component-specific (i.e., it moves on parallelly in each component when there are multiple components in a network). 
\begin{definition}~\cite{slikker07}
A Network game $v \in \mathbb H^N$ is zero monotonic if  $v(g')-v(L_i(g'))\geq v(g'-L_i(g'))$ for $g' \subseteq g$ and $i \in N(g')$.
\end{definition} 
\subsection{Recursive formula}
We now present a recursive formula for the weighted  position value. The formula is in the same spirit as the recursive formula for position value that is used by Slikker in \cite{slikker07}.
\begin{theorem}\label{theorem4}
Let $v$ be a component additive value function and $g$ be a network. Then for all $h \in C(g)$ and all $i \in N(h)$, we have, 
\begin{equation}
Y^{w}_i(g,v)=\frac{1}{|g|}\left[ \sum\limits_{l\in g_i}w^i_l[v(g)-v(g\backslash l)] +\sum\limits_{l\in g} Y^w_i(g\backslash l,v) \right].
\end{equation}
\end{theorem}
\begin{proof} Let, $h \in C(g)$ and $i \in N(h)$, then
\begin{align*}
	\sum\limits_{l\in g} \sum\limits_{g'\subseteq g} \frac{\sum\limits_{l'\in g_i} w^i_{l'}}{|g'|} \lambda_{g'}(v) 
	=&\sum\limits_{l\in g} \sum\limits_{g'\subseteq g:l\in g'} \frac{\sum\limits_{l'\in g_i} w^i_{l'}}{|g'|} \lambda_{g'}(v)
	+\sum\limits_{l\in g} \sum\limits_{g'\subseteq g:l\notin g'} \frac{\sum\limits_{l'\in g_i} w^i_{l'}}{|g'|} \lambda_{g'}(v)
\end{align*}
This would further imply that
\begin{align*}	
	& \sum\limits_{l\in g} \sum\limits_{g'\subseteq g} \frac{\sum\limits_{l'\in g_i} w^i_{l'}}{|g'|} \lambda_{g'}(v)
	-\sum\limits_{l\in g} \sum\limits_{g'\subseteq g:l\in g'} \frac{\sum\limits_{l'\in g_i} w^i_{l'}}{|g'|} \lambda_{g'}(v)
	=\sum\limits_{l\in g} \sum\limits_{g'\subseteq g\backslash l} \frac{\sum\limits_{l'\in g_i} w^i_{l'}}{|g'|} \lambda_{g'}(v)
\end{align*}
It follows that
\begin{align*}			
	& \sum\limits_{l\in g} \sum\limits_{g'\subseteq g} \frac{\sum\limits_{l'\in g_i} w^i_{l'}}{|g'|} \lambda_{g'}(v)
	-\sum\limits_{l\in g} \sum\limits_{g'\subseteq g:l\in g'} \frac{\sum\limits_{l'\in g_i} w^i_{l'}}{|g'|} \lambda_{g'}(v)
	=\sum\limits_{l\in g} \sum\limits_{g'\subseteq g\backslash l} \frac{\sum\limits_{l'\in g_i} w^i_{l'}}{|g'|} \lambda_{g'}(v)
\end{align*}
Therefore, we have
\begin{align*}	
 & |g| \sum\limits_{g'\subseteq g} \frac{\sum\limits_{l'\in g_i} w^i_{l'}}{|g'|} \lambda_{g'}(v) 
	-|g'| \sum\limits_{g'\subseteq g} \frac{\sum\limits_{l'\in g_i} w^i_{l'}}{|g'|} \lambda_{g'}(v)
	=\sum\limits_{l\in g} \sum\limits_{g'\subseteq g\backslash l} \frac{\sum\limits_{l'\in g_i} w^i_{l'}}{|g'|} \lambda_{g'}(v)\\
\end{align*}
Finally we have,
\begin{align*}	& |g-g'| \sum\limits_{g'\subseteq g} \frac{\sum\limits_{l'\in g_i} w^i_{l'}}{|g'|} \lambda_{g'}(v)
	=\sum\limits_{l\in g} \sum\limits_{g'\subseteq g\backslash l} \frac{\sum\limits_{l'\in g_i} w^i_{l'}}{|g'|} \lambda_{g'}(v)
\end{align*}
Now, 
\begin{align*}
	\frac{1}{|g|}\left[ \sum\limits_{l\in g_i}w^i_l[v(g)-v(g\backslash l)] +\sum\limits_{l\in g} Y^w_i(g\backslash l) \right]
	&= \frac{1}{|g|}\left[ \sum\limits_{l\in g_i}w^i_l \left(\sum\limits_{g'\subseteq g} \lambda_{g'}(v)\right) +\sum\limits_{l\in g} \sum\limits_{g'\subseteq g\backslash l} \frac{\sum\limits_{l'\in g_i} w^i_{l'}}{|g'|} \lambda_{g'}(v) \right]\\
	&= \frac{1}{|g|}\left[ \sum\limits_{g'\subseteq g} \sum\limits_{l\in g_i}w^i_l \lambda_{g'}(v) +|g-g'| \sum\limits_{g'\subseteq g} \frac{\sum\limits_{l'\in g_i} w^i_{l'}}{|g'|} \lambda_{g'}(v) \right] \\
	&= \frac{1}{|g|}\sum\limits_{g'\subseteq g} \left[ \sum\limits_{l\in g_i}w^i_l \lambda_{g'}(v) +|g-g'| \frac{\sum\limits_{l'\in g_i} w^i_{l'}}{|g'|} \lambda_{g'}(v) \right]\\
	&= \frac{1}{|g|}\sum\limits_{g'\subseteq g} \left[ |g'| \frac{\sum\limits_{l\in g_i}w^i_l}{|g'|} \lambda_{g'}(v) +|g-g'| \frac{\sum\limits_{l\in g_i} w^i_{l}}{|g'|} \lambda_{g'}(v) \right]\\
	&= \frac{1}{|g|}\sum\limits_{g'\subseteq g} \left[ |g| \frac{\sum\limits_{l\in g_i}w^i_l}{|g'|} \lambda_{g'}(v) \right]\\
	&=\sum\limits_{g'\subseteq g} \frac{\sum\limits_{l\in g_i}w^i_l}{|g'|} \lambda_{g'}(v)\\
	&= Y^w_i(g,v).
	\end{align*}
	\end{proof}
\subsection{The bidding process}
In this section, we introduce and analyze a mechanism that implements the weighted position value. The mechanism is defined recursively and component-wise for a Network game $(g,v)$. Let, $h \in C(g)$ be a component of $g$. 
If $|h|=0$, then the player $i$ in this component receives a stand-alone value of 0. Now, suppose $|h|=k \geq 1$ and the mechanism has been specified for components with at most $k-1$ links, the mechanism is played in multiple rounds. Each round consists of five stages (t=1-5) and after a round, the game either ends or new rounds start for the remaining components. In the following, we describe one such round in detail for a component $h \in C(g)$ with $k$ links.

 \begin{enumerate}
\item[\textbf{Stage 1.}] Each player $i \in N(h)$ makes a bid {$b^{i}_j \in \mathbb{R}$} for all $j \in N(h) \setminus i$. \\
Let {$$B^{i}=\sum_{j \in N(h)\setminus i}\sum_{l \subseteq g_i} w^{i}_lb_{j}^{i}-\sum_{j \in N(h)\setminus i}\sum_{l \subseteq g_j}w^{j}_lb_{i}^{j}$$} be the weighted net bid of player $i$ measuring its `relative' willingness to be the proposer. \\
Let $i^{*}$ be the player with the highest weighted net bid in this round. In case of a non-unique maximum value, we choose any of these maximal bidders to be the `winner' with equal probability. \\ 
 In the next stage $i^{*}$ the ``winner" becomes the proposer.  \textbf{Go to Stage 2}.
\item[\textbf{Stage 2.}] Every player $j \neq i^{*}$ divides the bid in his direction from player $i^{*}$ over the links of player $i^{*}$, i.e., she specifies $b^{i^{*}, l}_j \in \mathbb{R}$ for all $l \subseteq g_{i^{*}}$ under the condition that $$\sum_{l \subseteq g_{i^{*}}}b^{i^{*},l}_j=b^{i^{*}}_j.$$ Note here that $b^{i^{*}, l}_j$ is allowed to be negative.\footnote{The reason for this stage is that the bids in stage 1 are associated with a player, which can be identified with all his links. As only one link will be involved a division according to the weights of the links is required. This division being made by the other players will appear to be a driving force in all players being indifferent to the selection at stage 3.} \textbf{Go to Stage 3}.
\item[\textbf{Stage 3.}] Player $i^{*}$ chooses a link $l_{i^{*}} \in g_{i^{*}}$ and pays $b^{i^{*}, l}_j$ to every $j \in N(h) \setminus i^{*}$. \textbf{Go to stage 4}.
\item[\textbf{Stage 4.}] Player $i^{*}$ proposes payoffs $y_j$ to players $j \in N(h) \setminus i^{*}$. \textbf{Go to stage 5}.
\item[\textbf{Stage 5.}] At this stage, players other than $i^{*}$ sequentially accept or reject the proposed offers. As a result, two cases arise.
\begin{enumerate}

\item[\textbf{Case (a)}.] The offer is rejected by at least one of the players then the players of $N(h)$ proceed to play the next round where the set of links within $h$ is $|h\setminus l_{i^{*}}|.$ It is important to note that this next round consists of several separate (sub)-mechanisms in case $|h\setminus l_{i^{*}}|\geq 1.$ ~~~~~~\textbf{Stop}.

\item[\textbf{Case (b)}.] The offer is accepted. Then, each $j \in N(h) \setminus i^{*}$~ receives $y_{j}$, and player  $i^{*}$ obtains $$v(h)-\sum_{j \in N(h) \setminus i^{*}} y_{j}.$$
\\ We remark that these payoffs come on top of bids that were paid at stage 3 and, perhaps, payoffs from previous rounds. Hence, the final payoff  to player $j \in N(h) \setminus i^{*}$ is $$y_{j}+b_{j}^{i^{*},l}.$$ The final payoff  to player $i^{*}$ is $$v(h)-\sum_{j \in N(h) \setminus i^{*}}(y_{j}+b_{j}^{i^{*},l}).\;\;\;\;\;\textbf{ Stop}.$$ 

\end{enumerate}
\end{enumerate}
It is worth noting that, in this mechanism, decisions made in a particular stage of a specific round can incorporate payoffs that either include or exclude payoffs from previous stages and rounds, without affecting the overall analysis. For the sake of simplicity, we will not explicitly specify which specification we are referring to, but instead focus on comparing payoffs within a fixed specification.
\par The mechanism described above applied to situation $(g,v)$ and component $h \in C(g)$ will be denoted by $\Gamma (g,v).$

\begin{theorem}\label{theorem5}
For all {$g\in \mathbb G^N$}, $h\in C(g)$ and a zero monotonic  Network game $v \in \mathbb H^N$ there exists a sub-game perfect Nash equilibrium of the bidding mechanism such that the payoffs to the players coincide with the weighted position value.
\end{theorem}
\begin{proof}
Since the weighted position value is a generalization of the position value, so it is easy to see that the proof of this theorem follows similar reasoning as that of  Theorem 5.1 in \cite{slikker07}. However, we provide an independent proof in order to distinguish it from [20] so that the contributions of the weights in the model can be identified. Note that the difference in the implementation of the weighted position value to that of position value is in the way players make their respective bids to each other and it is due to the presence of the respective weights of the players. 
\\The proof follows by using induction on the number of links within a component. The theorem holds for networks and all their components with no links as well.
\par Let, $k\geq 1$, and let us assume that the theorem is true for all networks and all components with at most $k-1$ links. Let, $(g,v)$ be a Network game with $h \in C(g)$ and $|h|=k$. 
\\First, we show that the weighted position value is a sub-game perfect Nash equilibrium by constructing explicitly a sub-game perfect Nash equilibrium that yields the weighted position value. For that, we consider the strategies as follows:
\begin{enumerate}
\item[\textbf{Stage 1.}] Each player $i \in N(h)$ makes to any player $j \in N(h)\setminus i$ the bid
$$b_{j}^{i}=\sum_{l \subseteq g_i}[Y^{w}_j(g,v)-Y^{w}_j(g\setminus l,v)].$$
\item[\textbf{Stage 2.}]  Each player $j \in N(h)\setminus i^{*}$ chooses for all $l \subseteq g_{i^{*}}$\footnote{The last part of this expression is 0 in the node that follows the choices as specified in stage 1. However, this part is included since we cannot restrict our focus to the equilibrium path only to make sure that a strategy is a sub-game perfect Nash equilibrium.}:
\begin{equation}\label{eqn12}
b^{i^{*}, l}_j=[Y^{w}_j(g,v)-Y^{w}_j(g\setminus l,v)]+\sum_{l \subseteq g_{i^{*}}}w^{i^{*}}_l\big[ b^{i^{*}}_j-\sum_{l'\in g_{i^{*}}}[Y^{w}_j(g,v)-Y^{w}_j(g\setminus l',v)]\big].
\end{equation}
Here, $\sum_{l \subseteq g_{i^{*}}}b^{i^{*}, l}_j=b^{i^{*}}_j.$
\item[\textbf{Stage 3.}] Player $i^{*}$ makes a sub-game perfect choice by choosing a link $l$ that maximizes his payoff when taking into account the strategies at stage 4 and stage 5.
\item[\textbf{Stage 4.}] Player $i^{*}$ proposes to every player $j \in N(h)\setminus i^{*}$ payoff $$y^{i^{*}}_j=Y^{w}_j(g\setminus l_{i^{*}},v).$$ 
\item[\textbf{Stage 5.}] Player $j \in N(h)\setminus i^{*}$ accepts if $y^{i^{*}}_j\geq Y^{w}_j(g\setminus l_{i^{*}},v)$ and rejects the offer otherwise.
\\ Based on the integration of the aforementioned approaches, the game ends with the acceptance at stage 5. It can be readily verified that any player other than the proposer, i.e., $j \in N(h)\setminus i^{*}$acquires
\begin{align*}
y^{i^{*}}_j+b^{i^{*}, l}_j&=Y^{w}_j(g\setminus l_{i^{*}},v)+[Y^{w}_j(g,v)-Y^{w}_j(g\setminus l_{i^{*}},v)]\\
&=Y^{w}_j(g,v).
\end{align*}
The proposer $i^{*}$ acquires
\begin{align*}
v(h)-\sum_{j \in N(h)\setminus i^{*}}Y^{w}_j(g,v)&=v(h)-\sum_{j \in N(h)\setminus i^{*}}Y^{w}_j(g,v)\\
&=Y^{w}_{i^{*}}(g,v).
\end{align*}
It is important to note that the payoff determination is independent of the identity of the proposer $i^{*}$ and the identity of the link $l_{i^{*}}$. 
\par We now show that the strategies above result in weighted net bids equal to zero for all $i \in N$.
\begin{align*}
B^{i}&=\sum_{j \in N(h)\setminus i}\sum_{l \subseteq g_i}w^{i}_lb_{j}^{i}-\sum_{j \in N(h)\setminus i}\sum_{l \subseteq g_j}w^{j}_lb_{i}^{j}\\
&\sum_{j \in N(h)\setminus i}\big[\sum_{l \subseteq g_i}w^{i}_l[Y^{w}_j(g,v)-Y^{w}_j(g\setminus l,v)]-\sum_{l \subseteq g_j}w^{j}_l[Y^{w}_i(g,v)-Y^{w}_i(g\setminus l,v)]\big].
\end{align*}
By \textit{Weighted Balanced Link Contributions Property}, we have
 $$\sum\limits_{l\in g_j } w_l^j \Big(Y_i^{w}(g,v)-Y_i^{w}(g\backslash l,v)\Big)=\sum\limits_{l\in g_i } w_l^i \Big(Y_j^{w}(g,v)-Y_j^{w}(g\backslash l,v)\Big).$$
 Hence, $B^{i}=0.$ 
 \par It is now necessary to demonstrate that the aforementioned strategies form a sub-game perfect Nash equilibrium. It is easy to see that the strategies at stage 5 are the best responses. In case of rejection,
 all other players play the bidding mechanism with the link set $|h\setminus l_{i^{*}}|.$ By induction hypothesis, we have the weighted position value as the final outcome.
 \\To verify if the offers made by the proposer at stage 4 form a sub-game perfect Nash equilibrium, we examine the proposer's behavior in the event of their offer being rejected by the responder. It is seen that in case of rejection proposer $i^{*}$ obtains $$Y^{w}_{{i^{*}}}(g\setminus l_{i^{*}},v).$$ In case all others accept, due to the choices at stage 5 player $i^{*}$ can obtain at most $$v(h)-\sum_{j \in N(h)\setminus i^{*}}Y^{w}_j(g,v).$$ 
Since $v$ is zero-monotonic we have that $$v(h)\geq v(h\setminus l_{i^{*}}).$$ 
So,$$v(h)-\sum_{j \in N(h)\setminus i^{*}}Y^{w}_j(g,v)\geq Y^{w}_{{i^{*}}}(g\setminus l_{i^{*}},v) .$$
Hence, player $i^{*}$ maximizes his payoff by making offers as described in the strategy.
\par By definition, the strategies employed at stage 3 are sub-game perfect.
\par To verify if the strategies at stage 2 constitute the best response, we can observe that for any player $j$ in the set $N(h) \setminus i^{*}$, the choices made at stage 2 result in $j$'s payoff being independent of the decision made by player $i^{*}$ at stage 3, since
$$y_j+b^{i^{*}, l}_j=Y^{w}_j(g,v)+\sum_{l \subseteq g_{i^{*}}}w^{i^{*}}_l\big[ b^{i^{*}}_j-\sum_{l'\in g_{i^{*}}}[Y^{w}_j(g,v)-Y^{w}_j(g\setminus l',v)]\big].$$ 
As a result, player $i^{*}$ also has the flexibility to select any link at stage 3, since the same independence of payoffs holds for her. Let $\delta^{j}_l$ denote the change in $b^{i^{*}, l}_j$ for player $j$ when $b^{i^{*}, l}_j$ is updated to $b^{i^{*}, l}_j + \delta^{j}_l$, where $l$ is a fixed link in $g_{i^{*}}$.
\\When link $l$ is chosen at stage 3, it implies that player $j$ experiences a change in payoff of $\delta^{j}_l$, while player $i^{*}$ experiences a change of $-\delta^{j}_l$. As any change in player $j$'s strategy would result in a negative $\delta^{j}_l$ for at least one link $l$, player $i^{*}$ would select a link at stage 3 that increases her own payoff, since her payoff is independent of her choice. This would in turn decrease player $j$'s payoff.
\par Now, we examine the strategies at stage 1. Suppose player $i$ modifies her bids in a way that guarantees her becoming the proposer. In such a case, player $i$ would increase her total bids, but this would ultimately result in a decrease in her eventual payoff. On the other hand, if player $i$ adjusts her bids such that another player becomes the proposer with certainty, her payoff remains unchanged.
\\Lastly, consider the scenario where player $i$ has modified her bids and now has the maximum weighted net bid, while another player also has a weighted net bid equal to hers. In this case, there exist players $j$ and $k$ in the set $N(h)$ such that $b^{i}_j$ has been increased, while $b^{i}_k$ has been decreased due to the bid adjustments. To ensure that the weighted net bid of player $i$ is at least as much as the new weighted net bid of player $j$, player $i$ must have increased her total bid by at least the decrease in $b^{i}_j$. As a result, if player $i$ is chosen with a positive probability, she will increase her total bid and consequently decrease her eventual payoff. However, if player $i$ is not chosen with a positive probability, her payoff will remain unchanged.
This completes the proof.
\end{enumerate}
\end{proof}
The following theorem demonstrates that in any sub-game perfect Nash equilibrium, a player's payoff aligns with her weighted position value, even when the proposer and the link are randomly selected using random devices. Conversely, any sub-game perfect Nash equilibrium guarantees the realization of the weighted position value with certainty.
\begin{theorem}\label{theorem6}
For all {$g\in \mathbb G^N$}, $h\in C(g)$, a zero-monotonic  Network game $v \in \mathbb H^N$ and each sub-game perfect Nash equilibrium in $\Gamma(g,v)$ the payoffs to the players coincides with the weighted position value.
\end{theorem}
\begin{proof}
The proof that any sub-game perfect Nash equilibrium results in the weighted position value involves a series of claims, which have been already demonstrated in \cite{slikker07} for the position value. As the proofs for these claims are mostly similar in nature, our attention will be limited to demonstrating that the payoffs of each player align with their weighted position value within any sub-game perfect Nash equilibrium.
\begin{enumerate}
\item[\textbf{Claim 1.}] \textit{In any sub-game perfect Nash equilibrium at stage 5, players who are not the proposer will accept the proposal if and only if  $y^{i^{*}}_j>Y^{w}_j(g\setminus l_{i^{*}},v)$ for all $j \in N(h)\setminus i^{*}$.Additionally, if  $y^{i^{*}}_j<Y^{w}_j(g\setminus l_{i^{*}},v)$ for at least one player $j \in N(h)\setminus i^{*}$, then the proposal will be rejected.}

\item[\textbf{Claim 2.}]\textit{In the case where $v(h)>v(h\setminus l_{i^{*}})$, all sub-game perfect Nash equilibria of the game that start at stage 4 must satisfy the following specifications:}
\textit{\begin{itemize}
\item At stage 4, player $i^{*}$ proposes $y^{i^{*}}_j=Y^{w}_j(g\setminus l_{i^{*}},v)$ to all $j \in N(h)\setminus i^{*}$. 
\item At stage 5, every player $j \in N(h)\setminus i^{*}$ accepts any offer $y^{i^{*}}_j$ if $y^{i^{*}}_m=Y^{w}_m(g\setminus l_{i^{*}},v)$ for all $m \in N(h)\setminus i^{*}.$
\end{itemize}} 
\textit{If $v(h)=v(h\setminus l_{i^{*}})$, then any sub-game perfect Nash equilibrium that does not satisfy the specifications in the first part of this claim, satisfies these specifications:} 
\textit{\begin{itemize}
\item At stage 4, player $i^{*}$ proposes $y^{i^{*}}_j\leq Y^{w}_j(g\setminus l_{i^{*}},v)$ to some player $j \in N(h)\setminus i^{*}$. 
\item At stage 5, this offer is rejected by some player $j \in N(h)\setminus i^{*}$.
\end{itemize}} 
\textit{In all sub-game perfect Nash equilibrium of this sub-game the payoffs to the players are given by $[v(h)-v(h\setminus l_{i^{*}})]-\sum_{j \in N(h)\setminus i^{*}}b^{i^{*}, l_{i^{*}}}_j$ for player $i^{*}$ and $Y^{w}_j(g\setminus l_{i^{*}},v)+b^{i^{*}, l_{i^{*}}}_j$ for all $j \in N(h)\setminus i^{*}.$  }
\end{enumerate}
\item[\textbf{Claim 3.}]\textit{In any sub-game perfect Nash equilibrium of a game that starts at stage 2, the strategy of player $j \in N(h)\setminus i^{*}$ at stage 2 results in the same payoff for player $j$ regardless of the choice of player $i^{*}$ at stage 3. Hence, for all $l \subseteq g_{i^{*}}$, the payoff of player $j$ remains unchanged. So, we have, 
 $$b^{i^{*}, l}_j=[Y^{w}_j(g,v)-Y^{w}_j(g\setminus l,v)]+\sum_{l \subseteq g_{i^{*}}}w^{i^{*}}_l\big[ b^{i^{*}}_j-\sum_{l'\in g_{i^{*}}}[Y^{w}_j(g,v)-Y^{w}_j(g\setminus l',v)]\big].$$}
  \item[\textbf{Claim 4.}]\textit{In any sub-game perfect Nash equilibrium, each player is indifferent about the selection of proposer among $max\{B^{i}:i \in N(h)\}$}.
  \item[\textbf{Claim 5.}]\textit{In any any sub-game perfect Nash equilibrium, $B^{i}=0$ for all $i \in N(h)$}.
  \item[\textbf{Claim 6.}]\textit{In any sub-game perfect Nash equilibrium, the payoff of each of the players coincides with their weighted  position values.}
    \par Consider a sub-game perfect Nash equilibrium. Let $i\in N(h)$. If player $i$ is the proposer and she chooses link $l \subseteq g_i,$ then her payoff equals $$w^{i}_lx^{i,l}_i=Y^{w}_i(g\setminus l,v)+w^{i}_l[v(h)-v(h\setminus l)]-\sum_{l \subseteq g_i}w^{i}_l\sum_{j \in N(h)\setminus i}b^{i,l}_{j}.$$
    If $j \in N(h)\setminus i$ is the proposer and she proposes $l \subseteq g_j$ then player $i$ receives $$w^{j}_lx^{j,l}_i=Y^{w}_i(g\setminus l,v)+w^{j}_lb^{j,l}_{i}.$$
    Hence, the sum of the payoffs of player $i$ over all proposer-link combinations is given by
    \begin{align*}
    \sum_{j \in N(h)}\sum_{l \subseteq g_j}w^{j}_lx^{j,l}_i=&\sum_{l \subseteq g_i}\Big[Y^{w}_i(g\setminus l,v)+w^{i}_l[v(h)-v(h\setminus l)]-\sum_{l \subseteq g_i}w^{i}_l\sum_{j \in N(h)\setminus i}b^{i,l}_{j}\Big]+\\
    &\sum_{j \in N(h)\setminus i}\sum_{l \subseteq g_j}[Y^{w}_i(g\setminus l,v)+w^{j}_lb^{j,l}_{i}]\\
    &=\sum_{l \subseteq g_i}w^{i}_l(v(h)-v(h\setminus l)+\sum_{j \in N(h)}\sum_{l \subseteq g_j}Y^{w}_i(g\setminus l,v)-B^{i}\\
    &=\sum_{l \subseteq g_i}w^{i}_l[v(h)-v(h\setminus l)]+\sum_{l \subseteq g}Y^{w}_i(g\setminus l,v)\\
    &=|g|Y^{w}_i(g,v),
    \end{align*}
 where the third equality follows by Claim 5 and the last equality by Theorem \ref{theorem4}. Since, player $i$ is indifferent about the proposers(Claims 4 and 5)  and about the link chosen by the proposer (Claim 3) we have that all $x^{j,l}_i$ coincide, and hence, $x^{j,l}_i=Y^{w}_i(g,v)$ for all $j \in N(h)$ and $l \subseteq g_j.$ 
\end{proof}
 \section{Conclusions}
In this paper, we first provided two characterizations of the weighted position value for Network games. The formulation is in line with the position value characterized by \cite{Nouweland,Slikker05a}.  We also proposed a bidding mechanism for the class of weighted position values following a similar procedure. We showed that the weighted position value is supported by each sub-game perfect Nash equilibrium of the non-cooperative game stated in section 5. The bidding mechanism provides us with an explicit idea of the inherent interactions among the players. It helps us to understand the factors responsible for the existence of this particular class of allocation rules. Similar allocation rules can be modelled for games on hypergraphs. This we keep for our future study.


\end{document}